\documentclass[preprint,aps,showpacs,nofootinbib,preprintnumbers,amsmath,amssymb]{revtex4-1}
\usepackage{epsfig}
\usepackage{subfigure}
\usepackage{dcolumn}
\usepackage{bm}
\usepackage[usenames ,dvipsnames]{xcolor}
\usepackage{slashed}
\usepackage{graphicx,color}

\begin{document}
\title{Charmless two-body anti-triplet $b$-baryon decays}

\author{Y.K. Hsiao$^{1,2}$, Yu Yao$^{1}$, and C.Q. Geng$^{1,2,3}$}
\affiliation{
$^{1}$Chongqing University of Posts \& Telecommunications, Chongqing, 400065, China\\
$^{2}$Department of Physics, National Tsing Hua University, Hsinchu, Taiwan 300\\
$^{3}$Synergetic Innovation Center for Quantum Effects and Applications (SICQEA),\\
 Hunan Normal University, Changsha 410081, China
}
\date{\today}

\begin{abstract}
We study the charmless two-body decays of $b$-baryons $(\Lambda_b$, $\Xi_b^-$, $\Xi_b^0)$.
We find that ${\cal B}(\Xi_b^-\to \Lambda \rho^-)=(2.08^{+0.69}_{-0.51})\times 10^{-6}$ and
${\cal B}(\Xi_b^0\to \Sigma^+ M^-)=
(4.45^{+1.46}_{-1.09},11.49^{+3.8}_{-2.9},4.69^{+1.11}_{-0.79},2.98^{+0.76}_{-0.51})\times 10^{-6}$
for $M^-=(\pi^-,\rho^-,K^-,K^{*-})$, 
which are compatible to ${\cal B}(\Lambda_b\to p \pi^-,p K^-)$.
We also obtain that
${\cal B}(\Lambda_b\to \Lambda\omega)=(2.30\pm0.10)\times 10^{-6}$,
${\cal B}(\Xi_b^-\to\Xi^- \phi,\Xi^- \omega)\simeq
{\cal B}(\Xi_b^0\to\Xi^0 \phi,\Xi^0 \omega)=(5.35\pm0.41,3.65\pm0.16)\times 10^{-6}$
and
${\cal B}(\Xi^-_b\to\Xi^{-} \eta^{(\prime)})\simeq {\cal B}(\Xi^0_b\to \Xi^0 \eta^{(\prime)})=
(2.51^{+0.70}_{-0.46},2.99^{+1.16}_{-0.57})\times 10^{-6}$.
For the CP violating asymmetries, we show that
${\cal A}_{CP}(\Lambda_b\to p K^{*-})
={\cal A}_{CP}(\Xi_b^-\to \Sigma^0(\Lambda)K^{*-})=
{\cal A}_{CP}(\Xi_b^0\to \Sigma^+K^{*-})=(19.7\pm 1.4)\%$.
Similar to the charmless two-body $\Lambda_b$ decays, the $\Xi_b$ decays are accessible to the LHCb detector.
\end{abstract}

\maketitle
\section{introduction}
The charmful and charmless $\Lambda_b$ decays, such as
$\Lambda_b\to p M$~\cite{pdg,Aaltonen:2008hg},
$\Lambda_b\to \Lambda (\eta^{(\prime)},\phi)$~\cite{Aaij:2015eqa,Aaij:2016zhm},
$\Lambda_b\to D^-_s p,\Lambda_c^+ M$~\cite{Aaij:2014lpa}, and
$\Lambda_b\to p J/\psi M$~\cite{Pc_LHCb,LHCb1} with $M=(K^-,\pi^-)$,
have been measured by several experiments.
Recently, the LHCb Collaboration 
discovered the hidden-charm pentaquarks
in $\Lambda_b\to J/\psi p M$~\cite{Aaij:2015fea,Aaij:2016ymb},
and found the evidence of the time-reversal violating asymmetry
in $\Lambda_b\to p\pi^- \pi^+\pi^-$~\cite{Aaij:2016cla},
which indicates CP violation. Clearly,
the ${\cal B}_b$ decays are worthy of more theoretical and experimental studies,
where ${\cal B}_b$ denotes 
one of the anti-triplet $b$-baryons
of $\Lambda_b$, $\Xi_b^0$, and $\Xi_b^-$.
However, it seems more difficult to measure the $\Xi_b$ decays
due to $f_{\Xi_b}\simeq 1/10 f_{\Lambda_b}$
with $f_{{\cal B}_b}\equiv {\cal B}(b\to {\cal B}_b)$ as the fragmentation fraction.
To one's surprise,
apart from the charmful $\Lambda_b\to J/\psi \Lambda$ and
$\Xi_b^-\to J/\psi \Xi^-$ decays~\cite{pdg,Abazov:2007am,Aaltonen:2009ny},
the three-body $\Lambda_b$ and $\Xi_b$ modes
have been equally observed~\cite{Aaij:2014lpa,Aaij:2016nrq,Aaij:2016zab},
that is,
$\Lambda_b/ \Xi_b^0\to p \bar K^0 M$,
$\Lambda_b/\Xi_b^0\to \Lambda \pi^+\pi^-$,
$\Lambda_b/\Xi_b^0\to \Lambda K^+M$,
$\Xi_b^-\to p \bar K^- M$, and
$\Xi_b^-\to p \pi^-\pi^-$.

The charmless two-body $\Lambda_b$ decays
have been measured as follows~\cite{pdg,Aaltonen:2008hg,Aaij:2015eqa,Aaij:2016zhm}:
\begin{eqnarray}\label{data1}
&&{\cal B}(\Lambda_b\to p K^-,p\pi^-)=
(4.9\pm 0.9,4.1\pm 0.8)\times 10^{-6}\,,\nonumber\\
&&{\cal B}(\Lambda_b\to \Lambda \eta,\Lambda\eta^{\prime})=
(9.3^{+7.3}_{-5.3},<3.1)\times 10^{-6}\,,\nonumber\\
&&{\cal B}(\Lambda_b\to \Lambda \phi)=
(5.18\pm 1.04\pm 0.35^{+0.67}_{-0.62})\times 10^{-6}\,,
\end{eqnarray}
where $\Lambda_b\to \Lambda \phi$ can be viewed as the first observed vector mode,
while the results of 
$\Lambda_b\to \Lambda (\eta,\eta^{\prime})$ are still consistent with
the theoretical relation of
${\cal B}(\Lambda_b\to \Lambda \eta)\simeq
{\cal B}(\Lambda_b\to\Lambda\eta^{\prime})$~\cite{Ahmady:2003jz,Geng:2016gul}.
As the counterparts of the $\Lambda_b$ cases,
the two-body $\Xi_b$ decays of $\Xi_b^0\to \Sigma^+ M$, $\Xi_b^-\to \Lambda M$, and
$\Xi_b^{0,-}\to \Xi^{0,-}(\eta^{(\prime)},\phi)$ 
 should be explored experimentally, whereas no such decay has yet been observed.
Similar to the experimental situation, theoretically, 
even though the two-body $\Lambda_b$ decays have been well studied
in Refs.~\cite{Lu:2009cm,Wang:2013upa,Wei:2009np,Hsiao:2014mua,
Liu:2015qfa,Zhu:2016bra,Ahmady:2003jz,Geng:2016gul},
the $\Xi_b$ cases are barely explored 
except those in Refs.~\cite{Hsiao:2015txa,Hsiao:2015cda,He:2015fwa}.
In addition,
the CP-violating asymmetry (CPA) of ${\cal A}_{CP}(\Lambda_b\to p K^{*-})$
predicted to be 20\%~\cite{Hsiao:2014mua}
suggests that there can be large CPAs in the $\Xi_b$ processes 
due to the same anti-triplet hadronic structure.
Moreover, some of the charmless two-body decays of ${\cal B}_b\to{\cal B}_n M$
with M being $\pi^0$, $\eta^{(\prime)}$, $\phi$, $\rho^0$ and $\omega$ remain unexplored.
To compare with the future data, in this paper we systematically study
the charmless two-body ${\cal B}_b\to {\cal B}_n M$ decays with
${\cal B}_n$ being denoted as the baryon octet
and $M$ the pseudoscalar or vector meson.

\section{Formalism}

In terms of the effective Hamiltonian at the quark level,
the amplitudes of
the charmless two-body ${\cal B}_b\to {\cal B}_n M$ decays under the factorization approach
can be decomposed as
the matrix elements of the ${\cal B}_b\to {\cal B}_n$ baryon transitions along with
the vacuum to meson productions ($0\to M$).
In our classification, the first types of 
amplitudes with the unflavored mesons of $\pi^0,\rho^0,\omega$ and $\phi$
are given by~\cite{ali,Geng:2016gul}
\begin{eqnarray}\label{amp1a}
&&{\cal A}({\cal B}_b\to {\cal B}_n M)=
\frac{G_F}{\sqrt 2}\bigg\{\bigg[\alpha_2\langle M|(\bar uu)_{V-A}|0\rangle+
\alpha_3\langle M|(\bar uu+\bar dd)_{V-A}|0\rangle
+
\alpha_4\langle M|(\bar qq)_{V-A}|0\rangle\nonumber\\
&&+\alpha_5\langle M|(\bar uu+\bar dd)_{V+A}|0\rangle
+\alpha_9\langle M|(2\bar uu-\bar dd)_{V-A}|0\rangle\bigg]
\langle {\cal B}_n|(\bar qb)_{V-A}|{\cal B}_b\rangle\nonumber\\
&&+\alpha_6\langle M|(\bar qq)_{S+P}|0\rangle\langle
{\cal B}_n|(\bar qb)_{S-P}|\Lambda_b\rangle\bigg\}\,,\nonumber\\
&&{\cal A}({\cal B}_b\to {\cal B}_n \phi)=\frac{G_F}{\sqrt 2}
\bar \alpha_3
\langle \phi|(\bar ss)_V|0\rangle\langle {\cal B}_n|(\bar sb)_{V-A}|{\cal B}_b\rangle\,,
\end{eqnarray}
with $(\bar q_i q_j)_{V(A)}=\bar q_i\gamma_\mu(\gamma_5) q_j$ and
$(\bar q_i q_j)_{S(P)}=\bar q_i(\gamma_5) q_j$, where
$\alpha_2=V_{ub}V_{uq}^*\,a_2$,
$\alpha_3=-V_{tb}V_{tq}^*\,a_3$,
$\alpha_4=-V_{tb}V_{tq}^*\,a_4$,
$\alpha_5=-V_{tb}V_{tq}^*\,a_5$,
$\alpha_6=V_{tb}V_{tq}^*2a_6$,
$\alpha_9=-V_{tb}V_{tq}^*\,a_9/2$,
and $\bar \alpha_3=-V_{tb}V_{ts}^*\,(a_3+a_4+a_5-a_9/2-a_{10}/2)$.
%
In the generalized factorization approach~\cite{ali}, 
the color-singlet currents as in Eq.~(\ref{amp1a}) are kept for the vacuum to meson production 
and the ${\cal B}_b\to {\cal B}_n$ transition, such that one derives
the parameters $a_i \equiv c^{eff}_i+c^{eff}_{i\pm1}/N_c^{eff}$ for $i=$odd (even)
with the effective Wilson coefficients $c_i^{eff}$ and color number  $N_c^{eff}$.
On the other hand, the color-octet currents lead to the amplitudes of 
$\langle M {\cal B}_n|(\bar q_\alpha q'_\beta)(q''_\beta b_\alpha)|{\cal B}_b\rangle$
with $\alpha$ and $\beta$ the color indices, which are non-factorizable and disregarded. 
Nonetheless, by effectively shifting  $N_c^{eff}$ from 2 to $\infty$,
the non-factorizable contributions 
have been demonstrated to be well accounted~\cite{ali}. 
Note that $\Lambda_b\to \Lambda\phi$~\cite{Geng:2016gul}
with $a_{3,5}$ is estimated to have the large non-factorizable effect,
in which $N_c^{eff}$ is found to be around 2.
The relevant decays from the amplitudes in Eq.~(\ref{amp1a}) are
\begin{eqnarray}
&&\Lambda_b\to nM,\;
\Xi_b^{-,0}\to \Sigma^{-,0}M,\;\;\;\;\;\;\;\;\;\;\;\;\;\,\text{(for q=d)}\;\nonumber\\
&&\Lambda_b\to(\Lambda,\Sigma^0) M,\;
\Xi_b^{-,0}\to \Xi^{-,0}M,\;\;\;\;\;\text{(for q=s)}\;\nonumber\\
&&\Lambda_b\to(\Lambda,\Sigma^0) \phi,\;\Xi_b^{-,0}\to \Xi^{-,0}\phi\,,
\end{eqnarray}
with $M=(\pi^0,\rho^0,\omega)$.
The second types of 
amplitudes with the flavored mesons are given by~\cite{Hsiao:2014mua}
\begin{eqnarray}\label{amp2a}
&&{\cal A}({\cal B}_b\to {\cal B}_n M)\nonumber\\
&=&\frac{G_F}{\sqrt 2}
\bigg\{(\alpha_1+\alpha_4)\langle M|(\bar qu)_{V-A}|0\rangle
\langle {\cal B}_n|(\bar u b)_{V-A}|{\cal B}_b\rangle
+\alpha_6\langle M|(\bar q u)_{S+P}|0\rangle
\langle {\cal B}_n|(\bar ub)_{S-P}|{\cal B}_b\rangle\bigg\}\,,\nonumber\\
&&{\cal A}({\cal B}_b\to {\cal B}_n \bar K^{(*)0})\nonumber\\
&=&\frac{G_F}{\sqrt 2}
\bigg\{\alpha_4\langle \bar K^{(*)0}|(\bar sd)_{V-A}|0\rangle
\langle {\cal B}_n|(\bar d b)_{V-A}|{\cal B}_b\rangle
+\alpha_6\langle \bar K^{(*)0}|(\bar s d)_{S+P}|0\rangle
\langle {\cal B}_n|(\bar db)_{S-P}|{\cal B}_b\rangle\bigg\}\,,\nonumber\\
&&{\cal A}({\cal B}_b\to {\cal B}_n K^{(*)0})=-\frac{G_F}{\sqrt 2}
V_{tb}V_{td}^* a_4
\langle K^{(*)0}|(\bar ds)_{V-A}|0\rangle\langle {\cal B}_n|(\bar sb)_{V-A}|{\cal B}_b\rangle\,,
\end{eqnarray}
with
$\alpha_1=V_{ub}V_{uq}^*a_1$,
where the explicit decay modes are
\begin{eqnarray}
&&\Lambda_b\to pM,\;\Xi_b^-\to(\Lambda,\Sigma^0)M,\;\Xi_b^0\to \Sigma^+ M,\nonumber\\
&&\Lambda_b\to n\bar K^{(*)0},\;\Xi_b^{-,0}\to \Sigma^{-,0}\bar K^{(*)0},\nonumber\\
&&\Lambda_b\to(\Lambda,\Sigma^0) K^{(*)0},\;\Xi_b^{-,0}\to \Xi^{-,0}K^{(*)0},
\end{eqnarray}
with $M=(\pi^-,\rho^-)$ for $q=d$ and $M=K^{(*)-}$ for $q=s$.
With the mesons of $\eta^{(\prime)}$, the third types of 
amplitudes are given by~\cite{Geng:2016gul}
\begin{eqnarray}\label{amp3a}
{\cal A}({\cal B}_b\to {\cal B}_n \eta^{(\prime)})&=&
\frac{G_F}{\sqrt 2}\bigg\{
\bigg[\beta_2\langle \eta^{(\prime)}|(\bar q' q')_A|0\rangle+
\beta_3\langle \eta^{(\prime)}|(\bar s s)_A|0\rangle+
\beta_4\langle \eta^{(\prime)}|(\bar q q)_A|0\rangle\bigg]\nonumber\\
&\times&\langle {\cal B}_n|(\bar qb)_{V-A}|{\cal B}_b\rangle
+\beta_6\langle \eta^{(\prime)}|(\bar qq)_P|0\rangle
\langle{\cal B}_n|(\bar q b)_{S-P}|{\cal B}_b\rangle\bigg\}\,,
\end{eqnarray}
where $q'=u$ or $d$,
$\beta_2=-V_{ub}V_{uq}^*\,a_2+V_{tb}V_{tq}^*(2a_3-2a_5+a_9/2)$,
$\beta_3=V_{tb}V_{tq}^*(a_3-a_5-a_9/2)$,
$\beta_4=V_{tb}V_{tq}^*a_4$, and
$\beta_6=V_{tb}V_{tq}^*2a_6$. In Eq.~(\ref{amp3a}), 
the corresponding decays are
\begin{eqnarray}
&&\Lambda_b\to n\eta^{(\prime)},\;
\Xi_b^{-,0}\to \Sigma^{-,0}\eta^{(\prime)},\;\text{(for q=d)}\;\nonumber\\
&&\Lambda_b\to(\Lambda,\Sigma^0) \eta^{(\prime)},\;
\Xi_b^{-,0}\to \Xi^{-,0}\eta^{(\prime)}.\;\text{(for q=s)}\;
\end{eqnarray}

\begin{table}[t]
\caption{${\cal B}_b$ to ${\cal B}_n$ transition form factors.}\label{f1&g1}
\begin{tabular}{|c|c|}
\hline
$\langle {\cal B}_n|(\bar qb)|{\cal B}_b \rangle$ &F(0)\\\hline
$\langle p|(\bar ub)|\Lambda_b \rangle$&$\sqrt{3\over 2}C_{||}$\\
$\langle n|(\bar db)|\Lambda_b \rangle$&$\sqrt{3\over 2}C_{||}$\\
$\langle \Lambda|(\bar sb)|\Lambda_b \rangle$&$C_{||}$\\
$\langle \Sigma^0|(\bar sb)|\Lambda_b \rangle$&0\\
\hline
\end{tabular}
\begin{tabular}{|c|c|}
\hline
$\langle {\cal B}_n|(\bar qb)|{\cal B}_b \rangle$ &$F(0)$\\\hline
$\langle \Sigma^+|(\bar ub)|\Xi_b^0 \rangle$&$-\sqrt{3\over 2}C_{||}$\\
$\langle \Lambda|(\bar db)|\Xi_b^0 \rangle$&$-{1\over 2}C_{||}$\\
$\langle \Sigma^0|(\bar db)|\Xi_b^0 \rangle$&$\sqrt{3\over 4}C_{||}$\\
$\langle \Xi^0|(\bar sb)|\Xi_b^0 \rangle$&$-\sqrt{3\over 2}C_{||}$\\
\hline
\end{tabular}
\begin{tabular}{|c|c|}
\hline
$\langle {\cal B}_n|(\bar qb)|{\cal B}_b \rangle$ &$F(0)$\\\hline
$\langle \Sigma^-|(\bar db)|\Xi_b^- \rangle$&$\sqrt{3\over 2}C_{||}$\\
$\langle \Lambda|(\bar ub)|\Xi_b^- \rangle$&${1\over 2}C_{||}$\\
$\langle \Sigma^0|(\bar ub)|\Xi_b^- \rangle$&$-\sqrt{3\over 4} C_{||}$\\
$\langle \Xi^-|(\bar sb)|\Xi_b^- \rangle$&$\sqrt{3\over 2}C_{||}$
\\\hline
\end{tabular}
\end{table}
In Eqs.~(\ref{amp1a}), (\ref{amp2a}), and (\ref{amp3a}),
the matrix elements of the ${\cal B}_b\to{\cal B}_n$ transitions
can be presented as~\cite{Hsiao:2015txa,Hsiao:2015cda}
\begin{eqnarray}\label{b1}
\langle {\cal B}_{n}|(\bar q b)_{V-A}|{\cal B}_b\rangle&=&
\bar u_{{\cal B}_{n}}(f_1\gamma_\mu-g_1\gamma_\mu\gamma_5)u_{{\cal B}_b}\,,\nonumber\\
\langle {\cal B}_n|(\bar q b)_{S+P}|{\cal B}_b\rangle&=&
\bar u_{{\cal B}_{n}}(f_S+g_P\gamma_5)u_{{\cal B}_b}\,,
\end{eqnarray}
where $f_{1,S}$ and $g_{1,P}$ are the form factors.
Note that the parameterizations of the first matrix elements
safely ignore the terms of
$\bar u_{{\cal B}_{n}}\sigma_{\mu\nu}q^\nu(\gamma_5)u_{{\cal B}_b}$ and
$\bar u_{{\cal B}_{n}} q^\mu(\gamma_5)u_{{\cal B}_b}$ 
that flip the helicity of the spinor, whereas
the (axial)vector quark currents conserve the helicity.
In the equations of motion, $(f_S,g_P)$ are related to $(f_1,g_1)$ as
$f_S=(m_{{\cal B}_b}-m_{{\cal B}_n})/(m_{b}-m_{q})f_1$ and
$g_P=(m_{{\cal B}_b}+m_{{\cal B}_n})/(m_{b}+m_{q}) g_1$, respectively,
whose momentum dependences are given by~\cite{Hsiao:2014mua}
\begin{eqnarray}\label{b2}
f_1(q^2)=\frac{f_1(0)}{(1-q^2/m_{{\cal B}_b}^2)^2}\,,\;
g_1(q^2)=\frac{g_1(0)}{(1-q^2/m_{{\cal B}_b}^2)^2}\,.
\end{eqnarray}
The ${\cal B}_b\to {\cal B}_n$ transition form factors
for different decay modes can be related by
the $SU(3)$ flavor and $SU(2)$ spin symmetries~\cite{Hsiao:2015cda,Brodsky1},
resulting in the connection of $F(0)\equiv g_1(0)=f_1(0)$
and the relations given in Table~\ref{f1&g1},
where $C_{||}$ has been extracted from the data of
${\cal B}(\Lambda_b\to pK^-)$ and ${\cal B}(\Lambda_b\to p\pi^-)$~\cite{Hsiao:2014mua}.
For the meson productions, the matrix elements read
\begin{eqnarray}
&&\langle P|(\bar q_1 q_2)_A|0\rangle=- i f_Pq_\mu\,,\;
(m_{q_1}+m_{q_2})\langle P|(\bar q_1 q_2)_P|0\rangle=-i f_P{m_P^2}\,,\nonumber\\
&&\langle V|(\bar q_1 q_2)_V|0\rangle= m_V f_V\epsilon_\mu\,,
\end{eqnarray}
where $M=(P,V)$ are denoted as the pseudoscalar and vector mesons, 
respectively, and~\cite{Beneke:2002jn}
\begin{eqnarray}
&& \langle \eta^{(\prime)}|(\bar s s)_{A}|0\rangle=- i f^{s}_{\eta^{(\prime)}}q_\mu\,,
 \langle  \eta^{(\prime)}|(\bar q q)_{A}|0\rangle=- i \frac{f^{q}_{\eta^{(\prime)}}}{\sqrt{2}}q_\mu\,,\nonumber\\
&&2m_s\langle \eta^{(\prime)}|(\bar s s)_{P}|0\rangle=-i h^{s}_{\eta^{(\prime)}}\,,
2m_q\langle \eta^{(\prime)}|(\bar q q)_{P}|0\rangle=-i\frac{h^{q}_{\eta^{(\prime)}}}{\sqrt{2}} \,,
\end{eqnarray}
with
$(f_P,f_V,f^s_{\eta^{(\prime)}},f^q_{\eta^{(\prime)}},h^s_{\eta^{(\prime)}},h^q_{\eta^{(\prime)}})$
decay constants, $q_\mu(\epsilon_\mu)$
the four-momentum (-vector polarization), and $\bar q q=(\bar uu,\bar dd)$.
The direct CP-violating asymmetry is defined by
\begin{eqnarray}\label{eq_acp}
{\cal A}_{CP}({\cal B}_b\to {\cal B}_n M)\equiv
\frac{\Gamma({\cal B}_b\to {\cal B}_n M)-\Gamma({\cal \bar B}_b\to {\cal \bar B}_n \bar M)}
{\Gamma({\cal B}_b\to {\cal B}_n M)+\Gamma({\cal \bar B}_b\to {\cal \bar B}_n \bar M)}\,,
\end{eqnarray}
where
$\Gamma({\cal B}_b\to {\cal B}_n M)$ and
$\Gamma({\cal \bar B}_b\to {\cal \bar B}_n \bar M)$ are the decay widths from
the particle and antiparticle decays, respectively.

\section{Numerical Results and Discussions }
For our numerical analysis, we use
the CKM matrix elements in the Wolfenstein parameterization, given by~\cite{pdg}
\begin{eqnarray}\label{B1}
&&(V_{ub},V_{tb})=(A\lambda^3(\rho-i\eta),\,1)\,,\nonumber\\
&&(V_{ud},V_{td})=(1-\lambda^2/2,\,A\lambda^3)\,,\nonumber\\
&&(V_{us},V_{ts})=(\lambda,\,-A\lambda^2),
\end{eqnarray}
with $(\lambda,\,A,\,\rho,\,\eta)=(0.225,\,0.814,\,0.120\pm 0.022,\,0.362\pm 0.013)$.
The effective Wilson coefficients $c_i^{eff}$
are adopted as~\cite{ali}
\begin{eqnarray}
c^{eff}_1&=&1.168,\;\; c^{eff}_2=-0.365\,,\nonumber\\
10^4 \epsilon_1 c^{eff}_3&=&64.7+182.3 \epsilon_1\mp 20.2\eta - 92.6\rho +27.9\epsilon_2\nonumber\\
&&+i(44.2-16.2 \epsilon_1\mp 36.8\eta -108.6\rho + 64.4 \epsilon_2),\,\nonumber\\
10^4 \epsilon_1 c^{eff}_4&=&-194.1-329.8 \epsilon_1\pm 60.7\eta +277.8\rho -83.7\epsilon_2\nonumber\\
&&+i(-132.6+48.5 \epsilon_1\pm 110.4\eta +325.9\rho -193.3 \epsilon_2),\,\nonumber\\
10^4 \epsilon_1 c^{eff}_5&=&64.7+89.8 \epsilon_1\mp 20.2\eta - 92.6\rho +27.9\epsilon_2\nonumber\\
&&+i(44.2-16.2 \epsilon_1\mp 36.8\eta -108.6\rho + 64.4 \epsilon_2),\,\nonumber\\
10^4 \epsilon_1 c^{eff}_6&=&-194.1-466.7 \epsilon_1\pm 60.7\eta +277.8\rho -83.7\epsilon_2\nonumber\\
&&+i(-132.6+48.5 \epsilon_1\pm 110.4\eta +325.9\rho -193.3 \epsilon_2),\,\nonumber\\
10^4 \epsilon_1 c^{eff}_9&=&-3.0-109.5 \epsilon_1\pm 0.9\eta+4.3\rho-1.3\epsilon_2\nonumber\\
&&+i(-2.0\pm 1.7\eta +5.0\rho -3.0 \epsilon_2),\,\nonumber\\
10^4 c^{eff}_{10}&=&37.5,\,
\end{eqnarray}
for the $b\to d$ ($\bar b\to \bar d$) transition, and
\begin{eqnarray}
&&c^{eff}_1=1.168,\,c^{eff}_2=-0.365\,,\nonumber\\
&&10^4 c^{eff}_3=241.9\pm 3.2\eta + 1.4 \rho + i(31.3\mp 1.4\eta + 3.2\rho),\,\nonumber\\
&&10^4 c^{eff}_4=-508.7 \mp 9.6\eta - 4.2\rho+ i(-93.9 \pm 4.2\eta - 9.6\rho) ,\,\nonumber\\
&&10^4 c^{eff}_5=149.4\pm 3.2\eta + 1.4\rho + i(31.3\mp 1.4\eta + 3.2\rho),\,\nonumber\\
&&10^4 c^{eff}_6=-645.5 \mp 9.6\eta- 4.2\rho + i(-93.9\pm 4.2\eta - 9.6\rho) ,\,\nonumber\\
&&10^4 c^{eff}_9=-112.2 \mp 0.1\eta- 0.1\rho + i(-2.2\pm 0.1\eta - 0.1\rho) ,\,\nonumber\\
&&10^4 c^{eff}_{10}=37.5 ,\,
\end{eqnarray}
for the $b\to s$ ($\bar b\to \bar s$) transition,
where $\epsilon_1=(1-\rho)^2+\eta^2$ and $\epsilon_2=\rho^2+\eta^2$.
The meson decay constants are taken to be~\cite{pdg,Becirevic:2013bsa,Beneke:2002jn,Ball}
\begin{eqnarray}\label{B2}
&&(f_{\pi},f_{K},f_{\rho},f_{K^{*}},f_{\omega},f_{\phi})=(0.130,\,0.156,\,0.205,\,0.217,\,0.195,\,0.231)\,\text{GeV}\,,\nonumber\\
&&(f^q_{\eta},f^q_{\eta^\prime},f^s_{\eta},f^s_{\eta^\prime})=
(0.108,\,0.089,\,-0.111,\,0.136)\,\text{GeV}\,,\nonumber\\
&&(h^q_{\eta},h^q_{\eta^\prime},h^s_{\eta},h^s_{\eta^\prime})=(0.001,\,0.001,\,-0.055,\,0.068)\,\text{GeV}\,.
\end{eqnarray}
In addition, the extraction from the data gives
$|C_{||}|=0.111\pm 0.007$~\cite{Hsiao:2014mua,Hsiao:2015cda} in Table~\ref{f1&g1}.
Subsequently, we obtain
the branching ratios and direct CPAs for
the two-body charmless $\Lambda_b$, $\Xi_b^-$ and $\Xi_b^0$ decays,
shown in Tables~\ref{tab_Lb}, \ref{tab_Xibm} and \ref{tab_Xib0}, respectively.

For the $\Lambda_b$ decays,
it is interesting to note that all $\Lambda_b\to \Sigma^0 M$ decays,
such as $\Lambda_b\to \Sigma^0 (\pi^0,\eta^{(\prime)},\phi,\rho^0,\omega)$,
have zero branching ratios, which are not listed in Table~\ref{tab_Lb}.
This is due to
$\langle \Sigma^0|(\bar sb)|\Lambda_b \rangle=0$, where
the $b$ to $s$ transition currents transform
$\Lambda_b$ to $\Lambda=(ud-du)s$ that does not correlate to $\Sigma^0=(ud+du)s$.
It is clear that these nonexistent decays with ${\cal B}=0$
can test the theory based on the factorization approach.
To get the values of ${\cal B}(\Lambda_b\to p \pi^-,p\rho^-)$ in Table~\ref{tab_Lb},
we have used $a_1\simeq 1.0$ as the input in the amplitudes. In contrast, 
though being the tree-dominated modes also,
we take $a_2=0.18\pm 0.05$ ($N_c^{eff}\simeq 2$)~\cite{Hsiao:2015txa,Hsiao:2015cda}
to calculate the decays of $\Lambda_b\to n(\pi^0,\rho^0, \omega)$.
While $\langle \pi^0(\rho^0)|(\bar u u+\bar d d)|0\rangle=0$
with $(\pi^0,\rho^0)=u\bar u-d\bar d$ makes
the $\alpha_{3,5}$ terms disappear in the first amplitude in Eq.~(\ref{amp1a}),
one obtains
${\cal B}(\Lambda_b\to \Lambda\pi^0,\Lambda\rho^0)\simeq
{\cal O}(10^{-8}-10^{-7})$.
On the other hand, $\Lambda_b\to \Lambda\omega$
with $\omega=u\bar u+d\bar d$ enhances
its contribution from the $\alpha_{3,5}$ terms in Eq.~(\ref{amp1a}), resulting in
${\cal B}(\Lambda_b\to \Lambda\omega)>{\cal B}(\Lambda_b\to \Lambda\rho^0)$.
Note that ${\cal B}(\Lambda_b\to n \bar K^0,n\bar K^{*0})=
(4.61^{+1.48}_{-0.90},3.09^{+1.64}_{-0.81})\times 10^{-6}$
are as large as the counterparts of
${\cal B}(\Lambda_b\to p K^-,p K^{*-})
=(4.49^{+1.06}_{-0.76},2.86^{+0.73}_{-0.49})\times 10^{-6}$,
whereas
${\cal B}(\Lambda_b\to \Lambda K^0,\Lambda K^{*0})=
{\cal O}(10^{-8},10^{-7})$ are
mainly due to the CKM suppression of $|V_{td}/V_{ts}|=0.225$, respectively.

For the $\Xi_b$ decays, we obtain
\begin{eqnarray}
{\cal B}(\Xi_b^-\to \Lambda \pi^-,\Lambda \rho^-)&=&(0.80^{+0.26}_{-0.20},2.08^{+0.69}_{-0.51})\times 10^{-6}\,,\nonumber\\
{\cal B}(\Xi_b^-\to \Lambda K^{(*)-})&=&(0.85^{+0.20}_{-0.14},0.54^{+0.14}_{-0.09})\times 10^{-6}\,,\nonumber\\
{\cal B}(\Xi_b^0\to \Lambda \bar K^{(*)0})&=&(0.82^{+0.26}_{-0.16},0.54^{+0.29}_{-0.14})\times 10^{-6}\,.
\end{eqnarray}
By inputing the form factors of $F(0)^2=(3/4,1/4)C_{||}^2$
for the $\Xi_b^-\to \Sigma^0$ and $\Xi_b^-\to \Lambda$ transitions,
we get ${\cal B}(\Xi_b^-\to \Sigma^0 M^-)\simeq 3{\cal B}(\Xi_b^-\to \Lambda M^-)$
for $M^-=(\pi^-,\rho^-,K^{(*)-})$, respectively, indicating that 
the $\Sigma$ modes can be larger than the $\Lambda$ ones in the $\Xi_b$ decays.
Explicitly, we have 
\begin{eqnarray}
{\cal B}(\Xi_b^0\to \Sigma^+ \pi^-,\Sigma^+ \rho^-)
&=&(4.45^{+1.46}_{-1.09},11.49^{+3.8}_{-2.9})\times 10^{-6}\,,\nonumber\\
{\cal B}(\Xi_b^0\to \Sigma^+ K^{(*)-})
&=&(4.69^{+1.11}_{-0.79},2.98^{+0.76}_{-0.51})\times 10^{-6}\,,\nonumber\\
{\cal B}(\Xi_b^-\to \Sigma^- \bar K^{(*)0})
&=&(5.14^{+2.52}_{-1.70},3.43^{+1.81}_{-0.90})\times 10^{-6}\,,
\end{eqnarray}
where the relation of
${\cal B}(\Xi_b^0\to \Sigma^+ M^-)\simeq {\cal B}(\Lambda_b\to p M^-)$
with $M^-=(\pi^-,\rho^-,K^{(*)-})$ can be traced back to
the same amplitudes in Eq.~(\ref{amp2a}) with the identical inputing form factors.
On the other hand, with $F(0)^2=(3/4,3/2)C_{||}^2$ for
$\Xi_b^-\to \Sigma^0$ and $\Xi_b^{0}\to\Sigma^{+}$, 
we find
${\cal B}(\Xi_b^-\to \Sigma^0 K^{(*)-})\simeq {\cal B}(\Xi_b^0\to \Sigma^+ K^{(*)-})/2$ and
${\cal B}(\Xi_b^0\to \Sigma^0 \bar K^{(*)0})\simeq {\cal B}(\Xi_b^-\to \Sigma^- \bar K^{(*)0})/2$.
For the decays with $\eta^{(\prime)}$,
the branching fractions are given by
\begin{eqnarray}\label{N_eta}
{\cal B}(\Xi^-_b\to\Xi^{-} \eta^{(\prime)}) &=&
(2.67^{+0.74}_{-0.49},3.19^{+1.24}_{-0.61})\times 10^{-6}\,,
\nonumber\\
{\cal B}(\Xi^0_b\to \Xi^0 \eta^{(\prime)})&=&
(2.51^{+0.70}_{-0.46},2.99^{+1.16}_{-0.57})\times 10^{-6}\,,
\end{eqnarray}
with ${\cal B}(\Xi^-_b\to\Xi^{-} \eta^{(\prime)})\simeq {\cal B}(\Xi^0_b\to \Xi^0 \eta^{(\prime)})$
to obey the isospin symmetry.
Note that the branching ratios of these $\eta^{(\prime)}$ modes in Eq.~(\ref{N_eta})
are about 1.5 times larger than
${\cal B}(\Lambda_b\to\Lambda \eta^{(\prime)})$ (see Table~\ref{tab_Lb}).
As a result, the decays of $\Xi_b\to\Xi \eta^{(\prime)}$ are 
promising to be measured.

The $\Lambda_b\to \Lambda\phi$ decay is sensitive to $N_c^{eff}$ (see Table~\ref{tab_Lb}).
To explain the data in Eq.~(\ref{data1}),
we fix $N_c^{eff}=2$ to get ${\cal B}(\Lambda_b\to \Lambda\phi)=(3.42\pm 0.26)\times 10^{-6}$,
which implies the sizeable non-factorizable effects for
${\cal B}_b\to {\cal B}_n (\omega,\phi)$. 
Explicitly, we predict that
\begin{eqnarray}
{\cal B}(\Lambda_b\to \Lambda\omega)&=&(2.30\pm0.10)\times 10^{-6}\,,\nonumber\\
{\cal B}(\Xi_b^{-,0}\to\Xi^{-,0} \phi)&=&(5.70\pm0.43,5.35\pm0.41)\times 10^{-6}\,,\nonumber\\
{\cal B}(\Xi_b^{-,0}\to\Xi^{-,0} \omega)&=&(3.85\pm0.17,3.62\pm0.16)\times 10^{-6}\,,
\end{eqnarray}
which can be used to test the non-factorizable effects.

For the CPAs,
since the $\Lambda_b$ and $\Xi_b^{-,0}$ decays are associated with
the same amplitudes, we obtain
\begin{eqnarray}
&&{\cal A}_{CP}(M^-)\equiv {\cal A}_{CP}(\Lambda_b\to pM^-)
={\cal A}_{CP}(\Xi_b^-\to \Sigma^0(\Lambda)M^-)=
{\cal A}_{CP}(\Xi_b^0\to \Sigma^+M^-)\,,
\end{eqnarray}
where ${\cal A}_{CP}(M^-)=(-3.9\pm0.4,-3.8\pm0.4,
6.7\pm0.4,19.7\pm 1.4)\%$
for $M^-=(\pi^-,\rho^-,K^-,K^{*-})$, respectively.
Note that both uncertainties from the non-factorizable effects 
and form factors have been eliminated 
in Eq.~(\ref{eq_acp}) due to the ratios, leading to
small errors for the CPAs in Tables~\ref{tab_Lb} and \ref{tab_Xibm}. 
It is interesting to see that ${\cal A}_{CP}(K^{*-})$ is around 20\%,
which is large and should be measurable by the LHCb experiment.
We remark that 
the large non-factorizable effects in ${\cal B}_b\to {\cal B}_n (\omega,\phi)$
would flip the signs of uncertainties in the corresponding CPAs.

\section{Conclusions}
We have systematically examined 
all possible two-body ${\cal B}_b\to {\cal B}_n M$ decays
with ${\cal B}_b=(\Lambda_b,\Xi_b^-,\Xi_b^0)$,
${\cal B}_n=(p,n,\Lambda,\Xi^{-,0},\Sigma^{\pm,0})$ and
$M=(\pi^{-,0},K^{-,0},\bar K^0,\rho^{-,0},\omega,\phi,K^{*-,0},\bar K^{*0})$.
Explicitly, we have found that
${\cal B}(\Xi_b^-\to \Lambda \rho^-)=(2.08^{+0.69}_{-0.51})\times 10^{-6}$,
${\cal B}(\Xi_b^0\to \Sigma^+ M^-)=
(4.45^{+1.46}_{-1.09},11.49^{+3.8}_{-2.9},4.69^{+1.11}_{-0.79},2.98^{+0.76}_{-0.51})\times 10^{-6}$
for $M^-=(\pi^-,\rho^-,K^-,K^{*-})$, 
${\cal B}(\Lambda_b\to \Lambda\omega)=(2.30\pm0.10)\times 10^{-6}$,
${\cal B}(\Xi_b^-\to\Xi^- \phi,\Xi^- \omega)\simeq
{\cal B}(\Xi_b^0\to\Xi^0 \phi,\Xi^0 \omega)=(5.35\pm0.41,3.65\pm0.16)\times 10^{-6}$,
and
${\cal B}(\Xi^-_b\to\Xi^{-} \eta^{(\prime)})\simeq {\cal B}(\Xi^0_b\to \Xi^0 \eta^{(\prime)})=
(2.51^{+0.70}_{-0.46},2.99^{+1.16}_{-0.57})\times 10^{-6}$.
For CP violation, we have obtained
${\cal A}_{CP}(\Lambda_b\to p K^{*-})
={\cal A}_{CP}(\Xi_b^-\to \Sigma^0(\Lambda)K^{*-})=
{\cal A}_{CP}(\Xi_b^0\to \Sigma^+K^{*-})=(19.7\pm 1.4)\%$.
We urge to have some dedicated experiments to confirm 
these large CP asymmetries.
In sum, we have demonstrated that
most of the charmless two-body anti-triplet $b$-baryon decays 
are accessible to  the LHCb detector.

\section*{ACKNOWLEDGMENTS}
We would like to thank Dr. Eduardo Rodrigues for useful discussions.
This work was supported in part by National Center for Theoretical Sciences,
MoST (MoST-104-2112-M-007-003-MY3), and
National Science Foundation of China (11675030).

\newpage
\begin{table}[t]\caption{Two-body $\Lambda_b$ decays,
where the first two errors for $({\cal B}, {\cal A}_{CP})$ 
come from the non-factorizable effects
and CKM matrix elements, respectively, 
while the third error for $\cal B$ is due to the form factors.}\label{tab_Lb}
\begin{tabular}{|l|c|c|}
\hline
  ${\cal B}_{b}\to {\cal B}_{n} M$ & ${\cal B}\times 10^{6}$ & ${\cal A}_{CP}\times 10^2$\\\hline
  $\Lambda_{b}\rightarrow p \pi^{-}$ & $4.25^{+1.04}_{-0.48}\pm0.74\pm0.56$ & $-3.9^{+0.0}_{-0.0}\pm0.4$ \\
  $\Lambda_{b}\rightarrow p K^{-}$ & $4.49^{+0.84}_{-0.39}\pm 0.26\pm0.59$ & $6.7^{+0.3}_{-0.2}\pm0.3$\\
  $\Lambda_{b}\rightarrow  n \pi^{0}$ & $0.10^{+0.03}_{-0.03}\pm0.01\pm0.01$ &$8.0^{+1.2}_{-1.4}\pm 0.3$ \\
  $\Lambda_{b}\rightarrow n \bar{K}^{0}$ & $4.61^{+1.31}_{-0.58}\pm 0.31\pm0.61$ & $1.1^{+0.0}_{-0.0}\pm0.0$\\
  $\Lambda_{b}\rightarrow \Lambda \pi^{0}$ & $(3.4^{+0.8}_{-0.4}\pm0.1\pm0.4)\times 10^{-2}$ & $0.0^{+0.0}_{-0.0}\pm0.0$\\
  $ \Lambda_{b}\rightarrow \Lambda K^{0}$ & $(9.4^{+2.3}_{-3.8}\pm 0.4\pm1.3)\times 10^{-3}$ & $0.2^{+0.1}_{-0.0}\pm0.0$\\
  \hline
  $\Lambda_{b}\rightarrow p \rho^{-}$ & $11.03^{+2.72}_{-1.25}\pm1.97\pm1.46$ & $-3.8^{+0.0}_{-0.0}\pm 0.4$ \\
  $\Lambda_{b}\rightarrow p K^{*-}$ & $2.86^{+0.62}_{-0.29}\pm0.11\pm0.51$ & $19.7^{+0.4}_{-0.3}\pm 1.4$ \\
  $\Lambda_{b}\rightarrow  n \rho^{0}$ & $0.18^{+0.09}_{-0.09}\pm0.02\pm0.02$ & $14.0^{+1.8}_{-1.8}\pm 1.0$ \\
  $\Lambda_{b}\rightarrow  n \omega$ & $0.22^{+0.16}_{-0.10}\pm0.03\pm0.03$ & 
  $-18.2^{+24.4}_{-\,\,\,4.2}\pm 1.6$ \\
  $\Lambda_{b}\rightarrow  n \phi$ & $0.02^{+0.17}_{-0.02}\pm0.00\pm0.00$ & $-8.8^{+7.4}_{-5.1}\pm 0.3$ \\
  $\Lambda_{b}\rightarrow n \bar{K}^{*0}$ & $3.09^{+1.57}_{-0.67}\pm0.21\pm0.41$ & $1.3^{+0.1}_{-0.1}\pm 0.0$ \\
  $\Lambda_{b}\rightarrow \Lambda \rho^{0}$  & $(9.5^{+3.0}_{-1.3}\pm 0.4\pm1.3)\times 10^{-2}$ & $2.3^{+0.7}_{-0.8}\pm 0.2$ \\
  $\Lambda_{b}\rightarrow \Lambda \omega $ & $0.71^{+1.59}_{-0.70}\pm 0.04\pm0.09$ & $3.6^{+4.8}_{-4.0}\pm 0.2$ \\
  $\Lambda_{b}\rightarrow \Lambda \phi$ & $1.77^{+1.65}_{-1.68}\pm 0.12\pm0.23$ & $1.4^{+0.7}_{-0.1}\pm 0.1$ \\
  $\Lambda_{b}\rightarrow \Lambda K^{*0}$ & $(9.2^{+4.7}_{-2.0}\pm0.4\pm1.2)\times 10^{-2}$ & $1.3^{+0.1}_{-0.1}\pm0.0$ \\
  \hline
  $\Lambda_{b}\rightarrow n \eta $  & $(6.9^{+2.7}_{-2.4}\pm0.9\pm 0.9)\times 10^{-2}$ & $-16.8^{+2.1}_{-2.1}\pm1.3$ \\
  $\Lambda_{b}\rightarrow n \eta^{\prime}$  & $(4.2^{+1.8}_{-1.8}\pm0.6\pm0.6)\times 10^{-2}$ & $-15.7^{+4.0}_{-5.6}\pm1.3$ \\
  $\Lambda_{b}\rightarrow \Lambda \eta $  & $1.59^{+0.38}_{-0.17}\pm0.11\pm 0.21$ & $0.4^{+0.2}_{-0.2}\pm0.0$ \\
  $\Lambda_{b}\rightarrow \Lambda \eta^{\prime}$  & $1.90^{+0.68}_{-0.23}\pm0.13\pm0.25$ & $1.6^{+0.1}_{-0.1}\pm0.1$ \\
  \hline
\end{tabular}
\end{table}

\begin{table}\caption{Two-body $\Xi_b^-$ decays
with the error descriptions being the same as Table~\ref{tab_Lb}.}\label{tab_Xibm}
\begin{tabular}{|l|c|c |}
\hline
${\cal B}_{b}\to {\cal B}_{n} M$ & ${\cal B}\times 10^{6}$ & ${\cal A}_{CP}\times 10^2$\\\hline
$\Xi^{-}_{b}\rightarrow \Xi^{-} \pi^{0}$ & $(5. 7^{+1.3}_{-0.6}\pm0.2\pm0.7)\times 10^{-2}$ & $0.0^{+0.0}_{-0.0}\pm 0.0$ \\
    $\Xi^{-}_{b}\rightarrow \Sigma^{-} \pi^{0}$& $0.11^{+0.04}_{-0.04}\pm0.01\pm0.01$ &  $8.0^{+1.2}_{-1.4}\pm0.3$\\
  $\Xi^{-}_{b}\rightarrow \Sigma^{0} K^{-}$ & $2.50^{+0.47}_{-0.22}\pm 0.15\pm0.33$ & $6.7^{+0.3}_{-0.2}\pm0.3$\\
 $\Xi^{-}_{b}\rightarrow \Lambda K^{-}$ & $0.85^{+0.16}_{-0.07}\pm 0.05\pm0.11$ & $6.7^{+0.3}_{-0.2}\pm0.3$\\
  $\Xi^{-}_{b}\rightarrow \Sigma^{-} \bar{K}^{0}$ & $5.14^{+1.46}_{-0.64}\pm0.35\pm0.68$ & $1.1^{+0.0}_{-0.0}\pm0.0$\\
$\Xi^{-}_{b}\rightarrow \Sigma^{0} \pi^{-}$ & $2.37^{+0.58}_{-0.27}\pm0.41\pm0.31$ & $-3.9^{+0.0}_{-0.0}\pm0.4$ \\
$\Xi^{-}_{b}\rightarrow \Lambda \pi^{-}$ & $0.80^{+0.20}_{-0.09}\pm0.14\pm0.11$ & $-3.9^{+0.0}_{-0.0}\pm0.4$ \\
  $ \Xi^{-}_{b}\rightarrow \Xi^{-} K^{0}$ &$(1.6 ^{+0.3}_{-0.6}\pm0.1\pm0.2)\times 10^{-2}$ & $0.2^{+0.1}_{-0.0}\pm0.0$\\

\hline
$\Xi^{-}_{b}\rightarrow \Xi^{-} \rho^{0}$ & $0.16^{+0.05}_{-0.02}\pm 0.01\pm0.02$ & $2.3^{+0.7}_{-0.8}\pm 0.2$ \\
  $\Xi^{-}_{b}\rightarrow \Xi^{-} \omega$ & $1.18^{+2.67}_{-1.17}\pm 0.07\pm0.16$ & $3.6^{+4.8}_{-4.0}\pm 0.2$ \\
  $\Xi^{-}_{b}\rightarrow \Xi^{-} \phi$ & $2.95^{+2.75}_{-2.80}\pm 0.20\pm0.39$ & $1.4^{+0.7}_{-0.1}\pm 0.1$ \\
$\Xi^{-}_{b}\rightarrow \Sigma^{-} \rho^{0}$ & $0.20^{+0.10}_{-0.10}\pm 0.03\pm0.03$ & $14.0^{+1.8}_{-1.8}\pm 1.0$ \\
  $\Xi^{-}_{b}\rightarrow \Sigma^{-} \omega $ &$0.24^{+0.17}_{-0.11}\pm0.04\pm0.03$ & $-18.2^{+24.4}_{-\,\,\,4.2}\pm 1.6$ \\
  $\Xi^{-}_{b}\rightarrow \Sigma^{-} \phi$ &$0.02^{+0.19}_{-0.02}\pm0.00\pm0.00$ & $-8.8^{+7.4}_{-5.1}\pm0.3$\\
 $\Xi^{-}_{b}\rightarrow \Sigma^{0} K^{*-}$ & $1.59^{+0.34}_{-0.16}\pm 0.06\pm0.21$ &$19.7^{+0.4}_{-0.3}\pm1.4$ \\
 $\Xi^{-}_{b}\rightarrow \Lambda K^{*-}$ & $0.54^{+0.12}_{-0.05}\pm0.02\pm0.07$& $19.7^{+0.4}_{-0.3}\pm1.4$ \\
     $\Xi^{-}_{b}\rightarrow \Sigma^{-} \bar{K}^{*0}$ & $3.43^{+1.75}_{-0.74}\pm 0.23\pm0.45$ & $1.3^{+0.1}_{-0.1}\pm0.0$ \\
   $\Xi^{-}_{b}\rightarrow \Sigma^{0} \rho^{-}$ & $6.12^{+1.51}_{-0.69}\pm1.09\pm0.81$  & $-3.8^{+0.0}_{-0.0}\pm0.4$ \\
        $\Xi^{-}_{b}\rightarrow \Lambda \rho^{-}$ & $2.08^{+0.51}_{-0.23}\pm0.37\pm0.27$& $-3.8^{+0.0}_{-0.0}\pm0.4$ \\
   $\Xi^{-}_{b}\rightarrow \Xi^{-} K^{*0}$ & $0.15^{+0.08}_{-0.03}\pm0.01\pm0.02$  & $1.3^{+0.1}_{-0.1}\pm0.0$ \\
   \hline
   $\Xi^{-}_{b}\rightarrow \Xi^{-} \eta $ & $2.67^{+0.63}_{-0.29}\pm0.19\pm0.35$ & $0.4^{+0.2}_{-0.2}\pm0.0$ \\
  $\Xi^{-}_{b}\rightarrow \Xi^{-} \eta^{\prime}$ & $3.19^{+1.14}_{-0.38}\pm0.21\pm0.42$& $1.6^{+0.1}_{-0.1}\pm0.1$ \\
   $\Xi^{-}_{b}\rightarrow \Sigma^{-} \eta$ & $(7. 6^{+3.0}_{-2.7}\pm1.0\pm1.0)\times 10^{-2}$ & $-16.8^{+2.1}_{-2.1}\pm1.3$ \\
   $\Xi^{-}_{b}\rightarrow \Sigma^{-} \eta^{\prime}$ & $(4.7^{+2.0}_{-2.0}\pm0.6\pm0.6)\times 10^{-2}$ & $-15.7^{+4.0}_{-5.6}\pm1.3$\\
\hline
\end{tabular}
\end{table}

\begin{table}\caption{Two-body $\Xi_b^0$ decays
with the error descriptions being the same as Table~\ref{tab_Lb}.}\label{tab_Xib0}
\begin{tabular}{|l|c|c|}
\hline
${\cal B}_{b}\to {\cal B}_{n} M$
& ${\cal B}\times 10^{6}$ & ${\cal A}_{CP}\times 10^2$\\\hline
 $\Xi^{0}_{b}\rightarrow \Xi^{0} \pi^{0}$ & $(5.3^{+1.2}_{-0.6}\pm0.2\pm0.7)\times 10^{-2}$&  $0.0^{+0.0}_{-0.0}\pm 0.0$ \\
     $\Xi^{0}_{b}\rightarrow \Sigma^{0} \pi^{0}$&  $(5.1^{+1.8}_{-1.7}\pm0.5\pm0.6)\times 10^{-2}$ & $8.0^{+1.2}_{-1.4}\pm0.3$\\
$ \Xi^{0}_{b}\rightarrow \Lambda \pi^{0}$& $(1.7^{+0.6}_{-0.5}\pm0.1\pm0.2)\times 10^{-2}$ &$8.0^{+1.2}_{-1.4}\pm0.3$\\
  $\Xi^{0}_{b}\rightarrow \Sigma^{0} \bar{K}^{0}$ & $2.41^{+0.68}_{-0.30}\pm0.16\pm0.32$& $1.1^{+0.0}_{-0.0}\pm0.0$\\
    $\Xi^{0}_{b}\rightarrow \Lambda \bar{K}^{0}$ & $0.82^{+0.23}_{-0.10}\pm0.06\pm0.11$ & $1.1^{+0.0}_{-0.0}\pm0.0$\\
   $\Xi^{0}_{b}\rightarrow \Sigma^{+} K^{-}$ &$4.69^{+0.87}_{-0.41}\pm0.27\pm0.62$ & $6.7^{+0.3}_{-0.2}\pm0.3$\\
    $\Xi^{0}_{b}\rightarrow \Sigma^{+} \pi^{-}$ &$4.45^{+1.09}_{-0.50}\pm0.77\pm0.59$ & $-3.9^{+0.0}_{-0.0}\pm0.4$ \\
  $ \Xi^{0}_{b}\rightarrow \Xi^{0} K^{0}$ & $(1.5^{+0.4}_{-0.6}\pm0.1\pm0.2)\times 10^{-2}$ &  $0.2^{+0.1}_{-0.0}\pm0.0$\\
\hline
  $\Xi^{0}_{b}\rightarrow \Xi^{0} \rho^{0}$ & $0.15^{+0.05}_{-0.02}\pm 0.01\pm0.02$ & $2.3^{+0.7}_{-0.8}\pm 0.2$ \\
  $\Xi^{0}_{b}\rightarrow \Xi^{0} \omega $  & $1.11^{+2.51}_{-1.10}\pm 0.07\pm0.15$ & $3.6^{+4.8}_{-4.0}\pm 0.2$ \\
  $\Xi^{0}_{b}\rightarrow \Xi^{0} \phi$  & $2.77^{+2.58}_{-2.63}\pm0.19\pm0.37$ & $1.4^{+0.7}_{-0.1}\pm 0.1$ \\
$\Xi^{0}_{b}\rightarrow \Sigma^{0} \rho^{0}$ & $(9.5^{+4.6}_{-4.5}\pm1.3\pm1.3)\times 10^{-2}$ &$14.0^{+1.8}_{-1.8}\pm 1.0$ \\
  $\Xi^{0}_{b}\rightarrow \Sigma^{0} \omega$ & $0.11^{+0.08}_{-0.05}\pm0.02\pm0.01$&$-18.2^{+24.4}_{-\,\,\,4.2}\pm 1.6$ \\
  $\Xi^{0}_{b}\rightarrow \Sigma^{0} \phi$ &$(1.0^{+8.7}_{-0.8}\pm0.0\pm0.1)\times 10^{-2}$ & $-8.8^{+7.4}_{-5.1}\pm0.3$\\
$\Xi^{0}_{b}\rightarrow \Lambda \rho^{0}$ & $(3.2^{+1.6}_{-1.6}\pm0.4\pm0.4)\times 10^{-2}$
 &$14.0^{+1.8}_{-1.8}\pm 1.0$ \\
  $\Xi^{0}_{b}\rightarrow \Lambda \omega$ & $(3.8^{+2.8}_{-1.8}\pm0.6\pm0.5)\times 10^{-2}$ & 
  $-18.2^{+24.4}_{-\,\,\,4.2}\pm 1.6$ \\
  $\Xi^{0}_{b}\rightarrow \Lambda \phi$ & $(0.3^{ +3.0}_{-0.3}\pm0.0\pm0.0)\times 10^{-2}$ & $-8.8^{+7.4}_{-5.1}\pm0.3$\\
 $\Xi^{0}_{b}\rightarrow \Sigma^{0} \bar{K}^{*0}$ & $1.61^{+0.82}_{-0.35}\pm0.11\pm0.21$ & $1.3^{+0.1}_{-0.1}\pm0.0$ \\
 $\Xi^{0}_{b}\rightarrow \Lambda \bar{K}^{*0}$ & $0.54^{+0.28}_{-0.12}\pm0.04\pm0.07$ &  $1.3^{+0.1}_{-0.1}\pm0.0$ \\
 $\Xi^{0}_{b}\rightarrow \Sigma^{+} K^{*-}$ & $2.98^{+0.64}_{-0.30}\pm 0.11\pm0.39$ &   $19.7^{+0.4}_{-0.3}\pm1.4$ \\
   $\Xi^{0}_{b}\rightarrow \Sigma^{+} \rho^{-}$ & $11.49^{+2.83}_{-1.30}\pm2.05\pm1.52$ &  $-3.8^{+0.0}_{-0.0}\pm0.4$ \\
$\Xi^{0}_{b}\rightarrow \Xi^{0} K^{*0}$ & $0.14^{+0.07}_{-0.03}\pm0.01\pm0.02$ & $1.3^{+0.1}_{-0.1}\pm0.0$ \\
  \hline
   $\Xi^{0}_{b}\rightarrow \Xi^{0} \eta $ & $2.51^{+0.59}_{-0.27}\pm0.17\pm0.33$& $0.4^{+0.2}_{-0.2}\pm0.0$  \\
  $\Xi^{0}_{b}\rightarrow \Xi^{0} \eta^{\prime}$ &$2.99^{+1.07}_{-0.36}\pm0.20\pm0.40$& $1.6^{+0.1}_{-0.1}\pm0.1$ \\
$\Xi^{0}_{b}\rightarrow \Lambda \eta$ &$(1.2^{+0.5}_{-0.4}\pm0.2\pm0.2)\times 10^{-2}$ & $-16.8^{+2.1}_{-2.1}\pm1.3$ \\
   $\Xi^{0}_{b}\rightarrow \Lambda \eta^{\prime}$ & $(7.4^{+3.2}_{-3.1}\pm 1.0\pm1.0)\times 10^{-3}$  & $-15.7^{+4.0}_{-5.6}\pm1.3$ \\
   $\Xi^{0}_{b}\rightarrow \Sigma^{0} \eta$ & $(3.6^{+1.4}_{-1.3}\pm 0.5\pm0.5)\times 10^{-2}$& $-16.8^{+2.1}_{-2.1}\pm1.3$ \\
   $\Xi^{0}_{b}\rightarrow \Sigma^{0} \eta^{\prime}$ & $(2.2^{+0.9}_{-0.9}\pm 0.3\pm0.3)\times 10^{-2}$& $-15.7^{+4.0}_{-5.6}\pm1.3$\\
 \hline
\end{tabular}
\end{table}

\end{document}